\documentstyle[epsfig]{aipproc}

\begin{document}
\title{Jet Flavour Identification at\\
the CLIC Multi-TeV $e^+e^-$ Collider}

\author{Marco Battaglia}

\address{CERN, CH-1211 Geneva 23 Switzerland}

\maketitle

\begin{abstract}
Jet flavour identification in multi-TeV $e^+e^-$ collisions is expected to 
provide insights on new phenomena at scales beyond those probed by the LHC. 
The anticipated high track density and jet collimation represent a new 
challenge to jet tagging algorithms. A method, based on the sampling of 
the jet charged multiplicity, sensitive to the long decay length and large 
decay multiplicity of heavy flavour hadrons is proposed and the expected 
performances for the tagging of $e^+e^- \rightarrow b \bar b$ events at 
$\sqrt{s}$ = 3~TeV are discussed. 
\end{abstract}

\section{Introduction \label{sec:intro}}

The physics programme of future high energy linear colliders (LC), designed to 
deliver $e^+e^-$ collisions at centre-of mass energies $\sqrt{s}$ = 0.1-5~TeV 
with luminosities in excess to $10^{34}$cm$^{-2}$s$^{-1}$, largely relies on 
the ability to identify the flavour of final state fermions with high 
efficiency and purity. A first phase will be devoted to the accurate study 
of the Higgs profile, if a relatively light Higgs boson exists as suggested by
theory, electro-weak data and the signals reported at LEP-2, to the precise 
determination of the top mass and to complement the LHC program in searching 
for signals of new physics at the TeV mass scale. In a second phase, a 
multi-TeV LC is expected to break new grounds by studying in 
detail the properties of new physics established by the LHC or the lower 
energy LC and by exploring the mass scale beyond 10~TeV for new phenomena. 
The CLIC concept aims for a linear collider providing collisions at 
centre-of-mass energies beyond the TeV frontier with a luminosity of 
$10^{35}$cm$^{-2}$s$^{-1}$ at $\sqrt{s}$ = 3~TeV, based on the two-beam 
acceleration scheme~\cite{clic}.

If a new resonance has been detected at the TeV scale, either at the 
LHC or at a lower energy $e^+e^-$ collider, CLIC will be able to copiously 
produce it and provide with accurate determinations of its couplings to 
establish its nature. 
If no evidence for new particles has been obtained by the LHC and 
the LC, CLIC will probe the energy domain beyond 10~TeV by indirect searches, 
relying on accurate determinations of flavour-specific electro-weak data 
($\Gamma_{f \bar f}$, $A^{f \bar f}_{FB}$, $A_{LR}$) sensitive to virtual 
processes and vertex corrections from new particle contributions~\cite{mb}.

Tagging and anti-tagging of $t$, $b$ and $\tau$ final states will also be 
instrumental in refining the knowledge of the Higgs sector by isolating 
signals of heavy Higgs bosons, which occur in SM extensions such as 
Supersymmetry, and by 
testing the Higgs potential by measuring multiple Higgs production 
($ZHHH$ and $\nu_e \bar \nu_e HH(H)$) cross sections and of Supersymmetry. 
These analyses may 
require the high energy and luminosity of CLIC in order to access pair 
produced heavy bosons and to collect enough statistics in the low cross-section
processes testing the Higgs triple and quartic couplings. The signal final 
states are expected to be characterised by large $b$ jet multiplicity in 
processes such as $e^+e^- \rightarrow H^+H^- \rightarrow t \bar b \bar t b 
\rightarrow W^+ b \bar b W^- \bar b b$ and $e^+e^- \rightarrow \nu_e \bar 
\nu_e H H (H) \rightarrow \nu_e \bar \nu_e b \bar b b \bar b (b \bar b)$.

\section{The Vertex Tracker at CLIC}

Several solutions for the design and the silicon sensor technology have been 
proposed for the LC vertex tracker. These consist of either a multi-layered 
detector, based on CCD or CMOS sensor technology, with stand-alone tracking 
capabilities or a three-layered hybrid pixel detector.
At CLIC, the anticipated background from $e^+e^-$ pairs produced in the 
interaction of the colliding beams will limit the approach to the interaction
region to about 3.0~cm compared to $\simeq$~1.5~cm foreseen for the lower 
energy projects~\cite{backg}. This is compensated by the increase of the 
short-lived hadron decay length due to the larger boost. At $\sqrt{s}$ = 
3~TeV, the average decay distance of a $B$ hadron is 9.0~cm, in the two-jet 
process $e^+e^- \rightarrow b \bar b$, and 2.5~cm in multi-parton $e^+e^- 
\rightarrow H^+H^- \rightarrow t \bar b \bar t b \rightarrow W^+ b \bar b W^-
\bar b b$ decays. Due to the large boost and large hadronic multiplicity,
the local detector occupancy in $e^+e^- \rightarrow b \bar b$ events is 
expected to increase by almost a factor of ten, to $>1$ particle mm$^{-1}$
from $\sqrt{s}$ = 0.5~TeV to  $\sqrt{s}$ = 3.0~TeV.
This indicates the need to design a large vertex tracker based on
small area pixel sensors, able to accurately reconstruct the trajectories of 
secondary particles originating few tens of centimetres away from the beam 
interaction point and contained in highly collimated hadronic jets. There are 
other background issues relevant to the conceptual design of a vertex tracker 
for CLIC. These are the rate of $\gamma \gamma \rightarrow {\mathrm{hadrons}}$
events, estimated at 4.0~BX$^{-1}$, and the neutron flux, possibly of the
order of 10$^{10}$~1~MeV-equivalent~n~cm$^{-2}$~year$^{-1}$. The need to 
reduce the number of $\gamma \gamma$ events overlapped to a $e^+e^-$ 
interaction requires fast time stamping capabilities while the neutron induced
bulk damage has to be considered in terms of sensor efficiency reduction.

A vertex tracker consisting of seven concentric Si layers located from 3.0~cm
to 30~cm from the beam interaction point and based on hybrid pixel sensors,
with 20~ns time stamping and radiation hardness capabilities, demonstrated for 
their LHC applications has been considered in this study. The layer spacing 
has been chosen to optimally sample the heavy hadron decay length, resulting 
in a closer spacing for the innermost layers (see Figure~\ref{fig:event}).   
A model of this tracker has been implemented in GEANT; $e^+e^- \rightarrow 
b \bar b$ events and pair background have been passed through the GEANT 
simulation assuming a solenoidal field $B$ of 6~T. 
 
\section{$b$ Multiplicity Tag at CLIC \label{sec:btag}}

The kinematics in multi-TeV $e^+e^-$ collisions, suggests that the extensions 
of the reconstruction and tagging algorithms, pioneered at LEP and SLC and 
further developed for application at a lower energy LC, may need to be 
reconsidered. The high jet collimation and large $b$ decay distance pose 
significant challenges to the track pattern recognition and reconstruction that
may affect the accuracy of $b$ and $c$ identification by secondary vertex 
and impact parameter tagging. These are being addressed by studying tracking
performances for different designs for the main tracker~\cite{ariane}. 
At the same time it is interesting to explore new quark tagging techniques, 
that profit from the kinematics of multi-TeV $e^+e^-$ collisions.
A $b$~tagging algorithm based on the tag of the steps in particle multiplicity
originating from the heavy hadron decay along its flight path has been 
developed and tested here. The principle adopted stems from a technique 
developed for charm photo-production experiments~\cite{framm}. At a multi-TeV 
linear collider, such as CLIC, the signal of the production and decay of a $b$
or $c$ hadron can be obtained from an analysis of the number of hits recorded 
within a cone centred on the jet direction as a function of the radial 
position of the detector layer. In the present implementation of the 
algorithm, a cone of half-aperture angle $\psi$ = 40~mrad, optimised to 
maximise the sensitivity in presence of background and fragmentation 
particles, defines the area of interest (AoI) on each detector layer. 
The decay of a highly boosted 
short-lived hadron is characterised by a step in the number of hits recorded
in each AoI, corresponding to the additional charged multiplicity generated 
by the decay products. 
\vspace*{-0.25cm}
\begin{figure}[hb!]
\begin{center}
\begin{tabular}{l r}
\epsfig{file=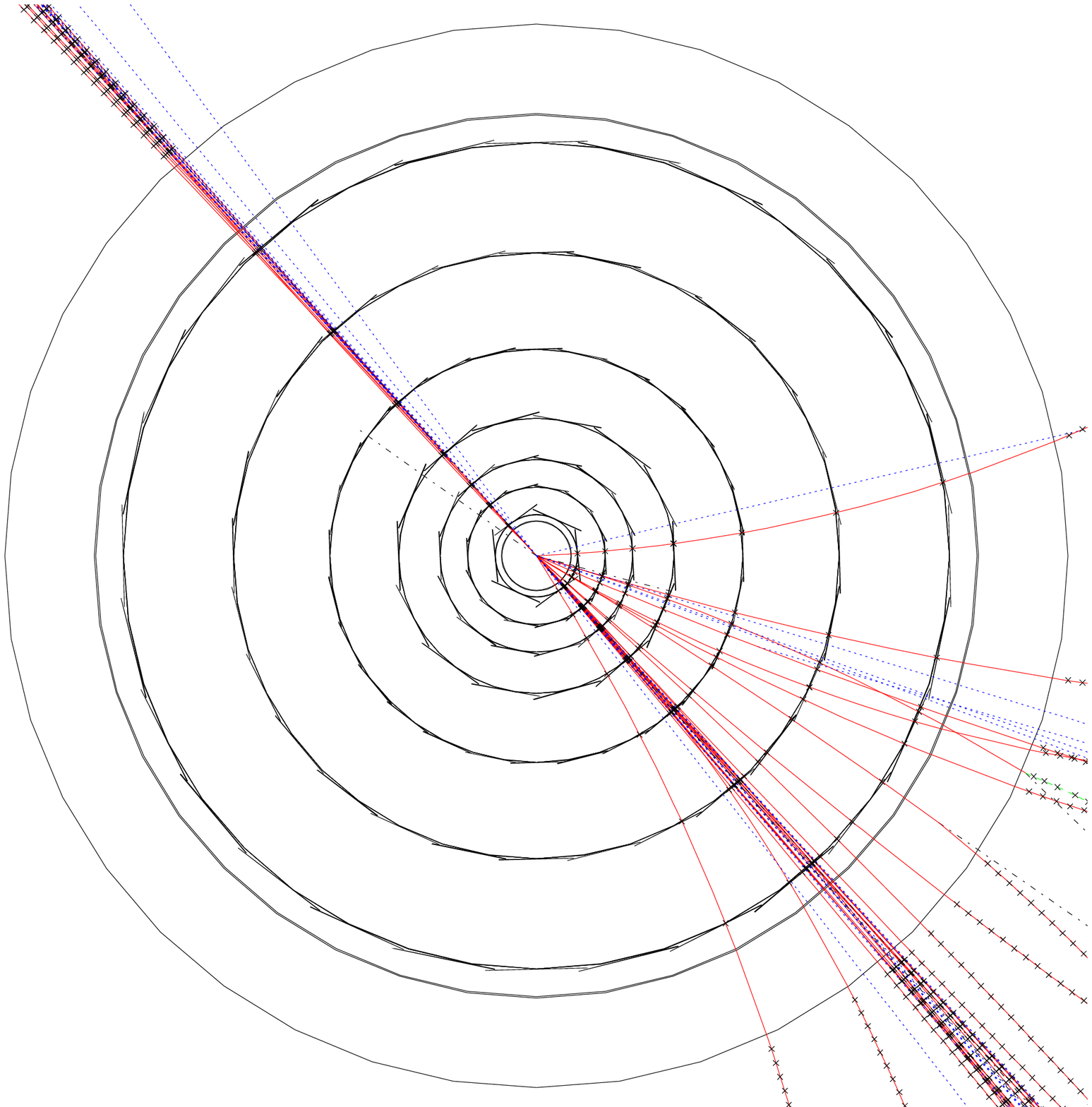,width=4.5cm,clip} & 
\epsfig{file=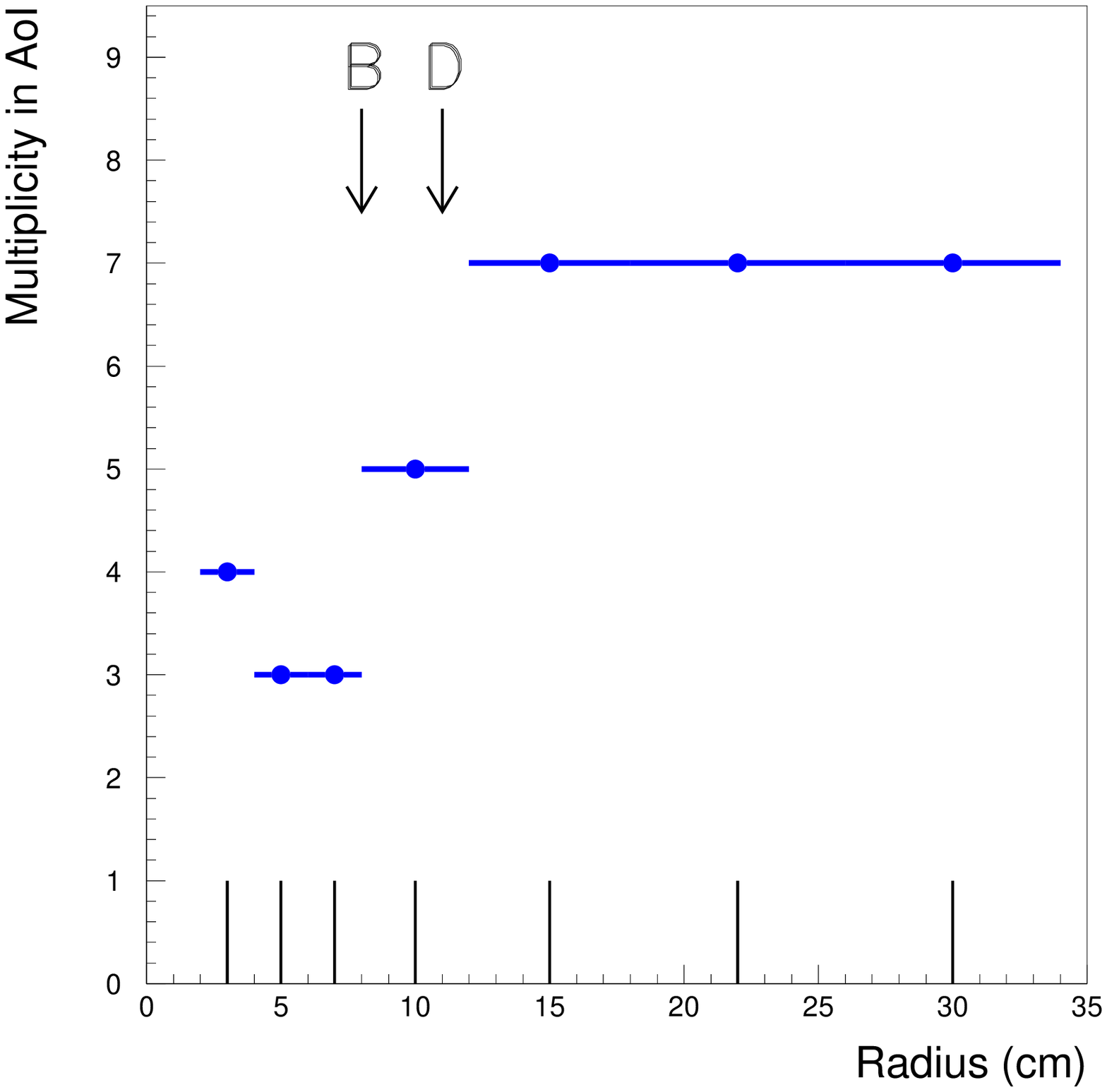,width=7.0cm,height=4.75cm,clip}\\
\end{tabular}
\caption{\label{fig:event} \sl Display of a $e^+e^- \rightarrow b \bar b$ event
at $\sqrt{s}$ = 3~TeV (left) with the detected hit multiplicity steps from the 
cascade decay of a long flying $B$ hadron (right).}
\end{center}
\end{figure}
\vspace*{-0.25cm}
Since the average charged decay multiplicity of a 
beauty hadron is about 5.2 \cite{bmult} and that of a charm hadron 
2.3, the $B$ decay signature can consists of either one or two steps, 
depending on whether the charm decay length exceeds the vertex tracker layer 
spacing. 
The number of background hits in the AoI is estimated to be 0.7, 
constant with the radius, the increase of the AoI surface being compensated by 
the background track density reduction due to the detector solenoidal field.
This background density is measured, by sampling the region outside the AoI, 
and subtracted. The effect of significant fluctuations of the number of these 
background hits and of low momentum curling tracks in the innermost layers can
be further removed by rejecting those jets where the hit multiplicity 
decreases with increasing radius or fluctuates by more than two units. The 
multiplicity pattern of a tagged $B$ decay is shown in Figure~\ref{fig:event}. 
\vspace*{-0.25cm}
\begin{figure}[ht!]
\begin{center}
\begin{tabular}{c c}
\epsfig{file=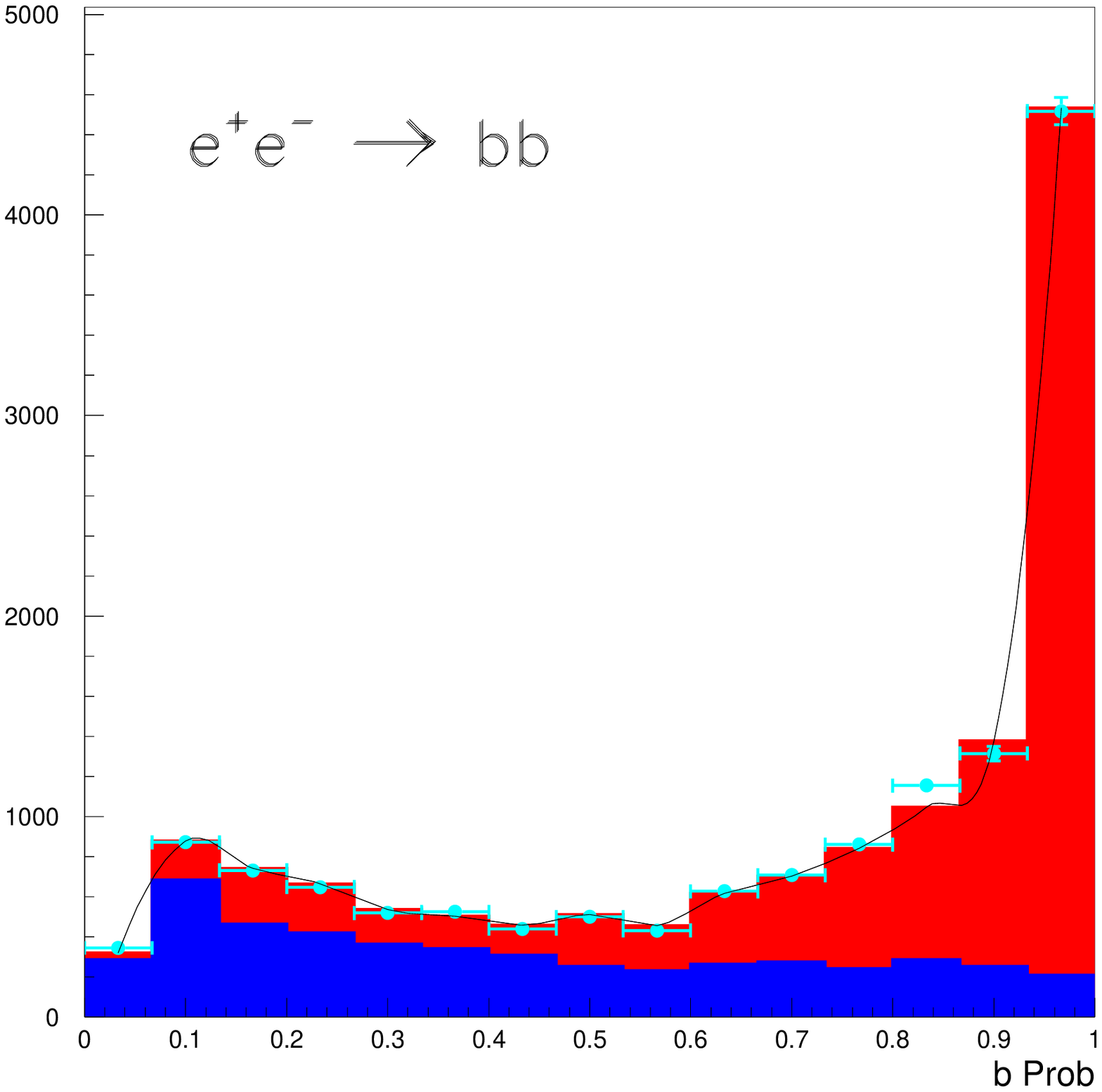,width=7.0cm,height=4.5cm} & 
\epsfig{file=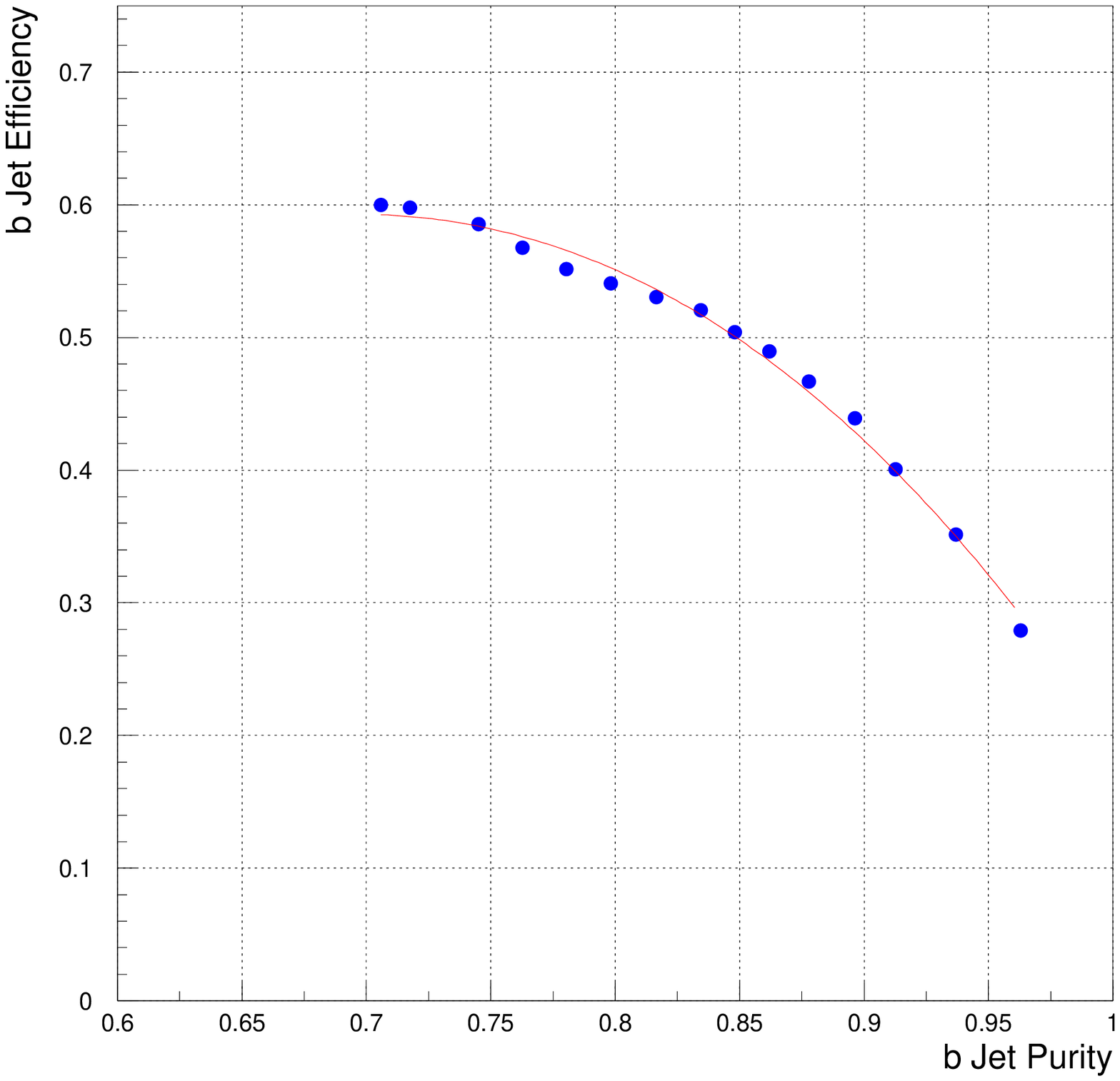,width=7.0cm,height=4.5cm}\\
\end{tabular}
\caption{\label{fig:tag} \sl The $b$ likelihood for jets in
$e^+e^- \rightarrow q \bar q$ events at $\sqrt{s}$ = 3~TeV (left) with a
multiplicity tag. The response for $b$ ($c$ and lighter) jets is shown in
light (dark) grey. The $b$-jet efficiency is given as a function of the purity
corresponding to different likelihood cut values (right).}
\end{center}
\end{figure}
\vspace*{-0.25cm}
Jets fulfilling the above criteria and with at least an upward multiplicity 
step larger than 1 and a total multiplicity increase larger than 2 have been 
considered. According to simulation 69\%, 29\% and
3\% of these jets are due to $b$, $c$ and light quarks respectively. The 
$b$ jets have been further discriminated using a $b$~likelihood based on the 
size, radial position and number of multiplicity steps, the fraction of the
jet energy and the invariant mass of the tracks originating at the detected 
multiplicity steps. The resulting likelihood for $b$ and lighter jets and the
$b$ efficiency and purity resulting from a cut on this discriminating variable
are shown in Figure~\ref{fig:tag}. By fitting that distribution for
the $b \bar b$ content, $\sigma(e^+e^- \rightarrow b \bar b)$ at 
$\sqrt{s}$ = 3~TeV can be determined to $\pm 0.01$~(stat) relative accuracy 
with an integrated luminosity of 1000~fb$^{-1}$, corresponding to one year of 
CLIC running at nominal luminosity.

I am grateful to A.~De Roeck, A.~Frey and R.~Settles for their comments and
suggestions.


\begin{references}

\bibitem{clic}
{\it A 3~TeV $e^+e^-$ Linear Collider Based on CLIC Technology}, 
G. Guignard (editor), CERN-2000-008.

\bibitem{mb}
M.~Battaglia, these proceedings.

\bibitem{backg}
D.~Schulte, these proceedings.

\bibitem{ariane}
A.~Frey, these proceedings.

\bibitem{framm}
A.~Albini {\it et al.} (NA1 Collaboration), {\it Phys. Lett.} {\bf B110}
(1982), 339.

\bibitem{bmult}
P.~Abreu {\it et al,} (DELPHI Collaboration), {\it Phys. Lett.} {\bf B425} 
(1998) 399;\\
G.~Brandenburg {\it et al.} (CLEO Collaboration), {\it Phys. Rev.} {\bf D61} 
(2000) 072002. 

\end{references}
\end{document}